# How graph neural network interatomic potentials extrapolate: role of the message-passing algorithm

Sungwoo Kang*

Computational Science Research Center, Korea Institute of Science and Technology (KIST), Seoul, 02792, Republic of Korea

*Email: sung.w.kang@kist.re.kr

**ABSTRACT**

Graph neural network interatomic potentials (GNN-IPs) are gaining significant attention due to their capability of learning from large datasets. Specifically, universal interatomic potentials based on GNN, usually trained with crystalline geometries, often exhibit remarkable extrapolative behavior towards untrained domains, such as surfaces or amorphous configurations. However, the origin of this extrapolation capability is not well understood. This work provides a theoretical explanation of how GNN-IPs extrapolate to untrained geometries. First, we demonstrate that GNN-IPs can capture non-local electrostatic interactions through the message-passing algorithm, as evidenced by tests on toy models and DFT data. We find that GNN-IP models, SevenNet and MACE, accurately predict electrostatic forces in untrained domains, indicating that they have learned the exact functional form of the Coulomb interaction. Based on these results, we suggest that the ability to learn non-local electrostatic interactions, coupled with the embedding nature of GNN-IPs, explains their extrapolation ability. We find that the universal GNN-IP, SevenNet-0, effectively infers non-local Coulomb interactions in untrained domains but fails to extrapolate the non-local forces arising from the kinetic term, which supports the suggested theory. Finally, we address the impact of hyperparameters on the extrapolation performance of universal potentials, such as SevenNet-0 and MACE-MP-0, and discuss the limitations of the extrapolation capabilities.

**I. INTRODUCTION**

Machine-learned interatomic potentials (MLIPs), which are machine learning models that predict the energies of given atomic configurations based on reference quantum mechanical calculation data, have gained much attention due to their speed, linear scalability, and high accuracy compared to density-functional theory (DFT) calculations.[1] Most early MLIP models predict the atomic energy of a given atom using input descriptors which represent local environments,[2–4] based on the assumption that non-local electrostatic interactions are effectively screened beyond a certain cutoff.[5] However, when it comes to systems involving large atomic charges, where electrostatic interactions are not screened within the cutoff radius, these conventional models fail to describe the potential energy surfaces (PES) of the system.[6] To overcome this issue, developments have been made to include non-local electrostatic interactions in conventional MLIPs.[7–13]

Recent developments in graph neural network interatomic potentials (GNN-IPs) have shown significant improvements in accuracy, data efficiency, and transferability compared to descriptor-based models.[14–17] These models incorporate a message-passing algorithm, suggesting their capability of extending the effective cutoff multiplied by the number of graph convolutions. However, it is still unclear whether these models effectively infer non-local electrostatics. Specifically, two previous studies explored this attribute of GNN-IPs and reached conflicting conclusions: Bochkarev et al. demonstrated that GNN-IPs, specifically ml-ACE and NequIP models, effectively train on the PES of molecules longer than the cutoff distance with considerable partial charges, indicating their ability to learn electrostatic interactions.[18] However, the charge distribution of atoms within a molecule is coupled with structural deformation, making it unclear whether GNN-IPs primarily learn PES based on structural deformations or they directly learn electrostatic interactions independently of geometric coupling. On the other hand, Nigam et al. showed that the message-passing algorithm can learn electrostatic interactions in random NaCl structures but concluded that this approach is much less effective compared to simply increasing the cutoff.[19] Specifically, they found that using the message-passing algorithm in linear models faces contradictory scenarios (see Fig. 1(a) for instance). However, it remains unknown how incorporating non-linearity would affect the results. For instance, including activation functions at nodes allows the information of atoms to be passed in a non-linear way, potentially capturing the correct electrostatic interactions regardless of atomic configurations. Overall, questions remain regarding the ability of GNN-IPs to model non-local electrostatic interactions effectively.

In addition to their ability to capture non-local interactions, GNN-IPs are attracting significant attention for their ability to learn a wide range of elemental systems. Utilizing this characteristic, they are often used to train universal interatomic potentials (UIPs) on massive databases (such as Materials Project[20]), enabling them to cover most elements in the periodic table.[21–27] These models demonstrate remarkable extrapolative capabilities in predicting properties across both untrained compositional and configurational spaces. As an example of compositional extrapolation, UIPs effectively predict energy values of configurations in untrained compositions, such as novel-composition materials resulting from stoichiometric substitutions.[24] In the case of configurational extrapolation, although UIPs are primarily trained on crystal structure databases, they show a remarkable ability to predict properties such as phonon dispersion, Li diffusivity, structural properties of amorphous phases, and even catalytic properties—though these predictions are not always highly accurate—for configurations not directly included in the training set.[22–25] It is suggested in ref. 24 that the use of atomic embedding is the origin of this extrapolation behavior because the knowledge learned from certain elemental systems can be transferred to other elemental systems. This transfer is possible because all elements share the same network structures and only differ by their embedding vectors. The embedding vectors are trainable, suggesting that the chemical relationships of elements are well represented in these vectors after training on a massive database. In this context, when visualizing the embedding vectors, it can be confirmed that elements with similar chemical properties are located close to each other.[24] Note that, in conventional descriptor-type models, this transfer is not feasible because the geometric information for elemental combinations is represented as separate descriptors, with distinct networks and weights used for different elements.[2,3] (We also note that recent developments have incorporated learnable embeddings in descriptor-based models.)[28,29] This embedding characteristic of GNN-IPs explains their ability to extrapolate untrained compositions within similar configurations. However, it is still unclear how GNN-IPs accurately extrapolate to untrained configurations, such as amorphous structural orders.[25]

In this study, we demonstrate that GNN-IPs effectively capture electrostatic interactions, which explains their extrapolation ability in untrained configurations. The ability of GNN-IPs to learn non-local interactions is confirmed using toy models consisting of randomly distributed charged particles. We find that the electrostatic interactions can be extrapolated within untrained ionic distances, indicating that GNN-IPs accurately learn electrostatic interactions. We also confirm that the electrostatic interactions and charge equilibrations can be described by GNN-IPs within DFT data, including the SrO crystal/disorder system and Au$_2$-adsorbed undoped/doped MgO slab. Finally, we provide a theoretical explanation of how GNN-IPs extrapolate to out-of-domain configurations using the extrapolative capability of non-local electrostatic interactions.

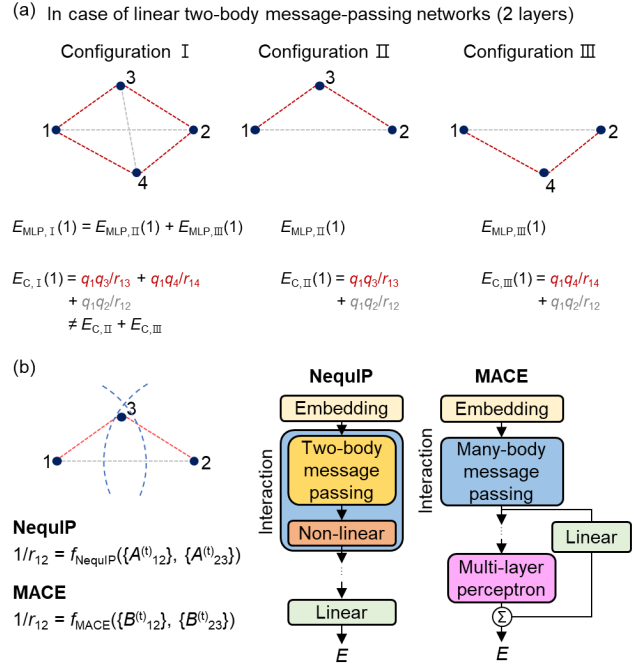

**FIG. 1.** (a) Schematic illustration of the message-passing algorithm for a graph neural network that consists of two convolution layers with two-body messages and linear activation functions. The distances between nodes 1 and 2 ($r_{12}$) and nodes 3 and 4 ($r_{34}$) are assumed to be larger than the cutoff, so message passing does not occur within these pairs. $E_{MLP}$ and $E_C$ stand for the energy calculated from the MLIP and the reference Coulomb energy, respectively. (b) Left: Schematic illustration describing the recombination process of predicting electrostatic interactions between two particle that are located at distance greater than the cutoff. Here, $A^{(t)}$ and $B^{(t)}$ represent the two-body and many-body features, respectively. Right: schematic illustration of the NequIP and MACE models. Two-body and many-body message passing include the additional functions such as self-interactions and ResNet-style updates.

## II. EXTRAPOLATION OF ELECTROSTATIC INTERACTIONS

### A. Capability of GNN-IPs of learning electrostatic interactions

We first discuss the necessity of non-linear GNN-IP models for describing electrostatic interactions, using the simple example illustrated in Fig. 1(a). We assume an extreme scenario to highlight the role of non-linear processes, where the GNN-IP model consists of two convolution layers with two-body messages and linear activation functions. In this case, when evaluating the atomic energy of atom 1, the contributions from atoms 3 and 4 are summed at atom 1 during the message-passing process, resulting in a simple linear sum, as shown in Fig. 1(a). The same argument applies to atoms 2-4. As a result, this leads to an unphysical relationship in the total energy for configurations I-III, highlighting the limitations of the model in representing all configurations simultaneously.

When non-linear processes are introduced into the model, this degeneracy can be avoided, as illustrated in Fig. 1(b). The right side of Fig. 1(b) shows the non-linearities in the two GNN-IP models, NequIP[15] and MACE,[14] used in this study. The NequIP model conducts interactions with two-body messages ($\{A^{(t)}\}$), where non-linear activation functions are applied to enhance the model's representability, similar to a multi-layer perceptron. In the MACE model, non-linearity is introduced through message-passing with many-body terms ($\{B^{(t)}\}$), which serve as a complete basis set.[30] The multi-layer perceptron for the final feature values also adds non-linearity. Thus, in these GNN-IP models, the unphysical degeneracy shown in Fig. 1(a) is avoided. In addition, the interaction between atoms 1 and 2 ($1/r_{12}$) can be inferred by recombining geometrical information from particles 1 and 3, and 2 and 3 or from particles 1 and 4, and 2 and 4:

$$\frac{1}{r_{12}} = \frac{1}{\sqrt{r_{13}^2 + r_{32}^2 + 2r_{13}r_{32}\cos(\theta_{132})}} = \frac{1}{\sqrt{r_{14}^2 + r_{42}^2 + 2r_{14}r_{42}\cos(\theta_{142})}}. \tag{1}$$

To represent Coulomb interactions beyond the cutoff, the out-of-cutoff distance, $r_{12}$, should be represented in terms of the in-cutoff distances and angles — namely $r_{13}$, $r_{14}$, $\theta_{132}$, and $\theta_{142}$ — via GNN-IP models. This information is included in the feature vector from the second convolution layer, as the filters in equivariant neural network contain both radial and angular components:

$$S_m^{(l)}(\vec{r}_{ij}) = R(r_{ij})Y_m^{(l)}(\hat{r}_{ij}), \qquad (2)$$

where $R(r_{ij})$ represents the radial component, generated from the multi-layer perceptron using basis embedding composed of radial Bessel functions and polynomial envelope functions, and $Y_m^{(l)}(\hat{r}_{ij})$ denotes the spherical harmonics. For the MACE model, it is guaranteed that out-of-cutoff distances can be described by in-cutoff distances and angles, if the degree of basis and the number of layers are sufficiently high, as the model expands the complete basis set. In the NequIP model, a self-interaction layer followed by an activation function within each convolution layer is identical to a single-layer perceptron. Therefore, if sufficient geometric information of out-of-cutoff particles is incorporated into the feature vector via message passing, any functional form can theoretically be represented, given a large enough number of features, according to the universal approximation theorem. Nevertheless, although a complete representation of out-of-cutoff distances is theoretically possible, achieving this within a practical model size (e.g., many-body order for MACE, number of layers for NequIP, and number of features for both models) is not guaranteed. In other words, the extent of the representability of these models—such as the impact of non-linear activations and message passing on electrostatic interactions—is not fully understood. In the following subsections, we will examine whether these models provide practical capabilities for describing electrostatic interactions in untrained domain.

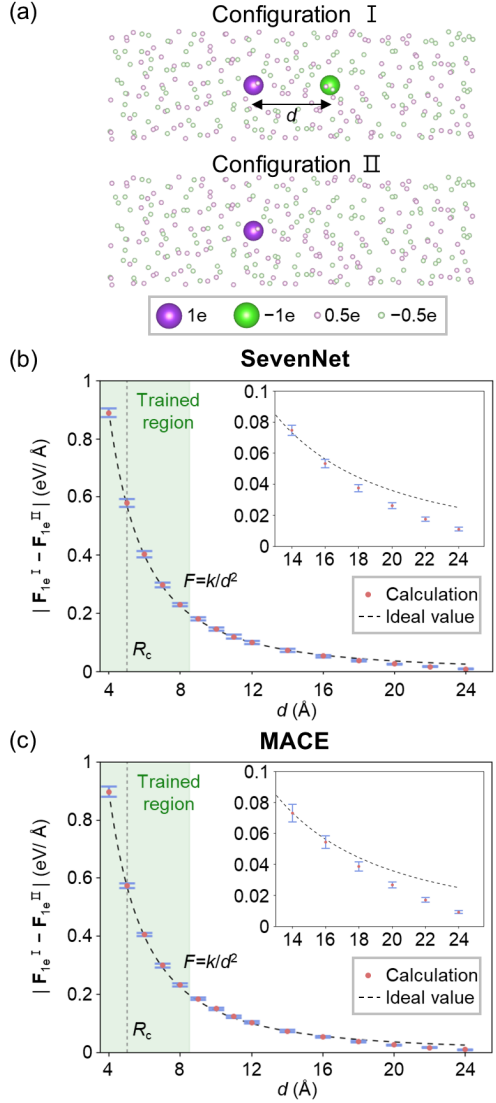

**FIG. 2.** (a) Schematic illustration of the toy model. Difference in forces acting on 1e particle in configurations I and II calculated by (b) SevenNet and (c) MACE with five convolution layers. Ideal value is plotted as a dotted line ($F=k/d^2$). Error bars represent the standard deviation of the test results. The trained regions are shaded in green. The vertical dotted line represents the cutoff ($R_c$) of the trained models.

**B. Toy model**

B.1. Toy model with charged medium particles

Initially, we construct a simple toy model comprising particles with point charges, applying only electrostatic interactions. Fig. 2(a) illustrates an example model used in this study. In the test simulations, we use pairs of structures (type I and II). In configuration type I, particles with charges of 1e, −1e, 0.5e, and −0.5e are included. The 1e and −1e particles are placed at a specified distance (*d*), while the 0.5e and −0.5e particles are placed randomly in a non-periodic cell. Configuration type II is identical to type I but omits the −1e particle in each cell. (Note that in the training process, we also include type III configurations, which contain −1e particles but omit 1e particles.)

We test two types of models: the SevenNet model[25] based on the NequIP architecture,[15] and the MACE model.[14] We set hyperparameters similar to those used in UIP models, such as SevenNet-0,[25] and MACE-MP-0[24] (see Table 1). While the size of the embedding vectors is set to 32, smaller than the 128 used in UIPs, this choice is reasonable given that our study focuses on four-element systems. The cutoff is set to 5 Å, and the number of convolutions is 5, making these models valid up to 25 Å. To assess their extrapolation ability, the models are trained at *d* of 4, 5, 6, 7, and 8 Å, and tested by evaluating the differences in forces acting on the 1e particles between configurations I and II ($\mathbf{F}_{1e}^{I}$ and $\mathbf{F}_{1e}^{II}$ respectively) at *d* ranging from 4 to 24 Å. Note that the ideal value corresponds to $k/d^2$ (*k*: Coulomb constant). Figs. 2(b),(c) show the results for SevenNet and MACE, respectively, both of which successfully predict $\mathbf{F}_{1e}^{I} - \mathbf{F}_{1e}^{II}$ in both the in-distribution and out-of-distribution regions. This extrapolation behavior indicates that both models accurately learn the functional form of Coulomb interactions.

**Table 1.** Hyperparameters used to train the toy models in this work, and those used for UIPs, SevenNet-0,[25] and MACE-MP-0.[24]

|  | SevenNet (This work) | SevenNet-0 | MACE (This work) | MACE-MP-0 |
|---|---|---|---|---|
| Number of convolution layers | 5 | 5 | 5 | 2 |
| Number of features | 32 | 128 (*l*=0), 64 (*l*=1), 32 (*l*=2) | 32 | 128 |
| $l_{max}$ | 3 | 2 | 3 | 3 |
| Correlation order | - | - | 3 | 3 |

Although the predictions agree with the reference points in most parts, we observe that the predictions slightly deviate above ~20 Å (the region of the last convolution layer) in both models, as shown in the insets of Figs 2(b),(c). To verify whether this error in the last convolution region appears with a different number of convolution layers, we perform the same test with the models with 3 convolution layers, as shown in Figs 3(a),(b) for SevenNet and MACE, respectively. We also observe that the accuracy of the model slightly decreases at distances of 13 and 14 Å. To systematically analyze this trend, we plot the relative mean error value ($\Delta \mathbf{F}_{ref} - \Delta \mathbf{F}_{mean}|/|\Delta \mathbf{F}_{ref}|$, where $\Delta \mathbf{F}_{mean}$ represents the mean of $\mathbf{F}_{1e}^{I} - \mathbf{F}_{1e}^{II}$ values) for SevenNet and MACE in Figs 3(c),(d), respectively. The normalized standard deviation (standard deviation divided by the absolute value of $|\mathbf{F}_{1e}^{I} - \mathbf{F}_{1e}^{II}|$) of both models is shown in Figs 3(e),(f). These standard deviation values can be interpreted as uncertainties in the prediction for long-range interactions, evaluated in the configurational ensemble that shares same geometric feature (in this case, *d*). The relative error and the normalized standard deviation show a similar trend in both models; these values increase as distance increases, especially in the last convolution layer region (3 layers: 10~15 Å, and 5 layers: 20~25 Å). Given that MACE employs general high-order-body interaction terms (in this study, up to four-body terms) yet exhibits similar error behavior to SevenNet, it can be inferred that the error in the last convolution layer region stems from the message-passing algorithm itself. Note that in the ACE framework,[30] including higher-order body terms do not affect model's performance, because Coulomb interactions are inherently two-body terms. However, when message passing is incorporated, the performance can vary with the body order of messages, as interactions beyond the cutoff are reconstructed using geometry information from within each cutoff. Therefore, incorporating higher-order terms can improve the model's ability to reconstruct interactions beyond cutoff by utilizing these terms as a basis set.

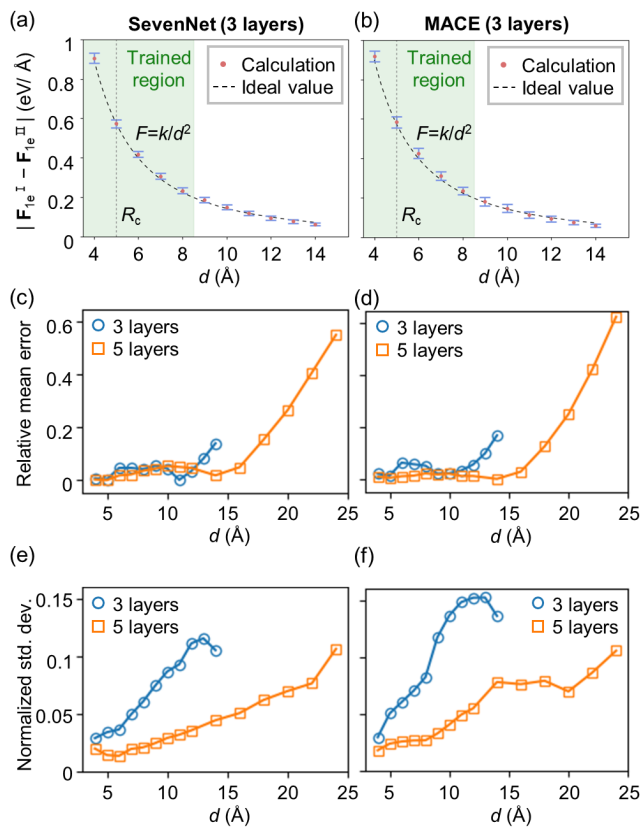

**FIG. 3.** Difference in forces acting on a 1e particle in configurations I and II calculated by (a) SevenNet and (b) MACE with three convolution layers. Relative mean error ($|\Delta \mathbf{F}_{ref} - \Delta \mathbf{F}_{mean}|/|\Delta \mathbf{F}_{ref}|$, where $\Delta \mathbf{F}_{mean}$ represents the mean of $\mathbf{F}_{1e}^{I} - \mathbf{F}_{1e}^{II}$ values) calculated by (c) SevenNet and (d) MACE as a function of $d$. Normalized standard deviation (standard deviation divided the absolute value of $|\mathbf{F}_{1e}^{I} - \mathbf{F}_{1e}^{II}|$) calculated by (e) SevenNet and (f) MACE as a function of $d$.

The varying accuracy in the different convolution regions can be rationalized as follows: As shown in Fig. 1(b), if the interactions between two particles beyond the cutoff are described through message passing, it should be recombined based on the geometrical relation between each particle with each of medium particles. Such recombination requires non-linear processes, such as the non-linear activation functions in each convolution block in NequIP or many-body expansions in each layer and the multi-layer perceptron in the final layer of MACE. In this context, the message-passing algorithm does not only extend the effective cutoff but also enhances the non-linear processes involved in capturing interactions. For instance, in a NequIP model with $N$ convolution layers, the first cutoff region undergoes $N$ activations, the second cutoff region undergoes $N-1$ activations, and the last cutoff region undergoes only one activation. Similarly, in a MACE model, the body-order associated with interactions between atoms in the first shell is larger than that representing interactions between an atom in the first cutoff region and an atom in the last convolution region, with a difference up to $NM$ (where $M$ is the maximum body order). As a result, interactions involving more message passing (father distance) have lower representability, leading to higher errors in electrostatic energies in the final activation layers for both the SevenNet (NequIP-based) and MACE models. The above claim is further supported by the fact that standard deviation of the 3-layer model is higher than that of the 5-layer model at the same $d$ (see Fig. 3), as atoms in the same region pass through more convolution layers in the 5-layer model, enhancing the representational capacity of each node. However, when considering the final convolution layer, the error is larger in the 5-layer model than in the 3-layer model. This is because, in the last layer, pairwise interactions are processed through only one convolution layer in both models, but the 5-layer model incorporates information from the larger number of nodes (atoms) during message passing than the 3-layer model. Consequently, the 5-layer model requires more parameters to effectively recombine the pairwise electrostatic interactions in the final convolution-layer distance.

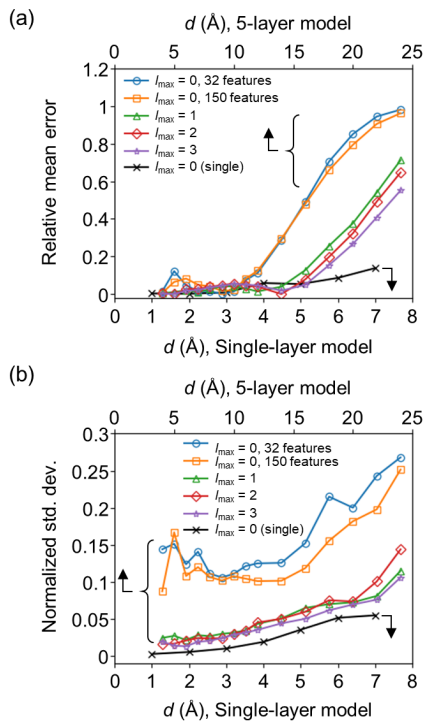

**FIG. 4.** (a) Relative mean error ($|\Delta \mathbf{F}_{ref} - \Delta \mathbf{F}_{mean}|/|\Delta \mathbf{F}_{ref}|$, where $\Delta \mathbf{F}_{mean}$ represents the mean of $\mathbf{F}_{1e}^{I} - \mathbf{F}_{1e}^{II}$ values), and (b) normalized standard deviations of SevenNet models with varying $l_{max}$, as a function of $d$. The models are all consists of 5 convolution layers with 5 Å cutoff, trained with training sets of $d$ = 4-8 Å. The number of features is set at 32 for $l_{max}$ = 1,2,3. The results of the single-layer SevenNet model are also presented. For comparison, the data are presented on the same relative scale, corresponding to the effective cutoff (defined as the cutoff multiplied by the number of convolutions).

We also note that errors and standard deviations do not increase linearly with an increasing $d$ in Fig. 3. In the case of MACE, the degree of basis terms decreases from the first to the last convolution layer. When the basis degree is sufficient to represent the desired function, further increasing degree does not significantly reduce the error values. This is why the error remains consistently low across most of the model but increases at the last convolution layer in the 3-layer model and at the second-to-last convolution layer in the 5-layer model. For NequIP, the number of non-linear functions applied decreases from the first to the last convolution region. The representational capacity of a multi-layer perceptron increases exponentially with the number of hidden layers.[31] Therefore, the error in the NequIP model does not decrease linearly with the number of involving convolutions.

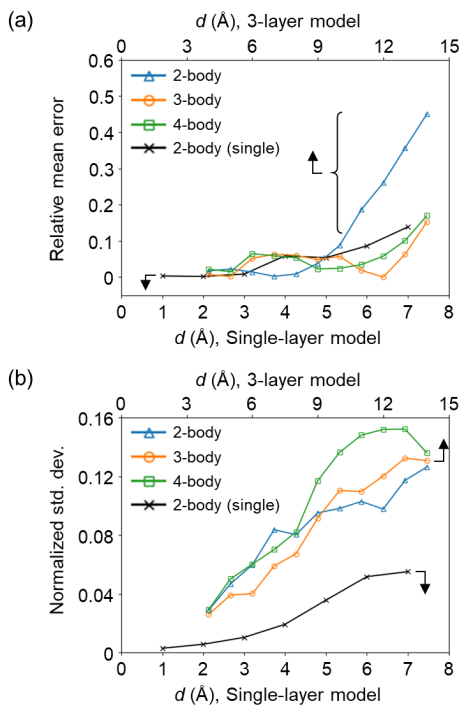

**FIG. 5.** (a) Relative mean error ($|\Delta \mathbf{F}_{ref} - \Delta \mathbf{F}_{mean}|/|\Delta \mathbf{F}_{ref}|$, where $\Delta \mathbf{F}_{mean}$ represents the mean of $\mathbf{F}_{1e}^{I} - \mathbf{F}_{1e}^{II}$ values), and (b) normalized standard deviations of MACE models with varying correlation order (body order) and the number of features, as a function of $d$. The models are all consists of 3 convolution layers with 5 Å cutoff, trained with training sets of $d$ = 4-8 Å. The results of the single-layer SevenNet model are also presented. For comparison, the data are presented on the same relative scale, corresponding to the effective cutoff (defined as the cutoff multiplied by the number of convolutions).

To further support that regions undergoing more convolutions tend to exhibit lower accuracy, we conduct the same numerical experiment using the SevenNet model with a single convolution layer and compare the results with those from the multi-layer model. Figs. 4,5 illustrate the relative standard deviations and relative errors of the 5-layer SevenNet models and 3-layer MACE models with diverse hyperparameters, respectively, compared to those of the single-layer SevenNet model with an increased cutoff (8 Å). The results indicate that the error and standard deviation of the single-layer models are lower in the extrapolation region compared to the multi-layer models in Fig. 2, which aligns with the findings in ref. 19. This is because the single-layer model explicitly incorporates two-body radial terms, eliminating the need for complex recombining processes in the message-passing algorithm, as explained in the earlier subsection. Therefore, this further supports the idea that the error and standard deviation observed in Fig. 3 are caused by the message-passing algorithm, rather than from an insufficient description of the local environment of the message terms. However, increasing the cutoff is significantly less efficient in terms of computational cost and memory compared to the message-passing algorithm.[32] Therefore, it is important to optimize the hyperparameters of message-passing algorithms to balance both accuracy for long-range interactions and computational efficiency.

In this sense, we analyze the impact of hyperparameters on the extrapolation performance of GNN-IP models. As shown in Table 1, we adopt computationally demanding settings similar to those used in UIP models. However, the question remains whether smaller GNN-IP models that focused on specific systems, can also effectively extrapolate electrostatic interactions. Fig. 4 includes the effect of $l_{max}$ on the extrapolation performance of the SevenNet model in predicting point-charge interactions. The results indicate that when $l_{max} = 0$, the networks exhibit high errors and large standard deviations, even with an increased number of features, 150. In contrast, networks with $l_{max}$ = 1, 2, or 3 show similar results, suggesting that increasing $l_{max}$ beyond 1 does not significantly improve performance. Fig. 5 includes the impact of correlation order on the extrapolation performance of the MACE model in predicting point-charge interactions. The extrapolation performance of the two-body model (correlation order = 1) is notably worse in the last convolution layer region than that of models considering interactions up to three-body (correlation order = 2) and four-body (correlation order = 3) terms.

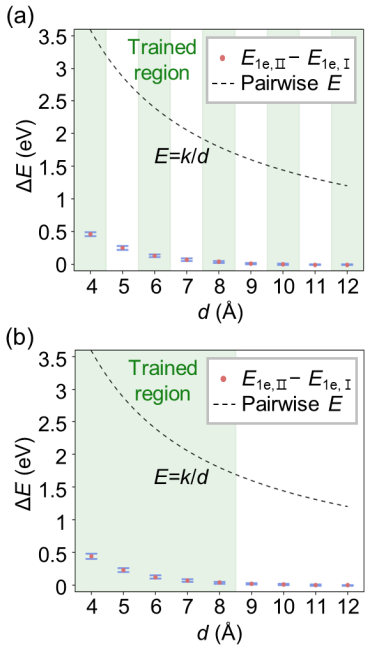

**FIG. 6.** Difference in atomic energy of 1e particle in configuration I and II for (a) interpolation test set and (b), extrapolation test set. The absolute value of pairwise electrostatic energy between 1e and −1e ($E=k/d$) is plotted in the dotted line. Error bars represent the standard deviations of the test results. The calculations are performed using SevenNet.

B.2. Atomic energy mapping

To gain insight into how GNN-IPs learn electrostatic interactions, we identify how electrostatic energies are stored in the form of atomic energy. Figs 6(a),(b) show the atomic energy differences for the 1e particle between configuration I and II ($E_{1e}^{I}$ and $E_{1e}^{II}$, respectively) for interpolation and extrapolation test sets, respectively. (The interpolation model is trained at $d$ of 4, 6, 8, 10, and 12 Å.) The dotted line plots $k/d$, corresponding to the pairwise electrostatic energy exerted on the 1e particle by the −1e particle. The values of $E_{1e}^{I} - E_{1e}^{II}$ values are significantly smaller than the pair-wise electrostatic energy, which indicates that electrostatic energy between an atom pair is distributed among neighboring atoms rather than being localized solely within the two atoms involved.

One might assume that this scenario arises from the ad hoc atomic energy mapping, where atomic energies are inaccurately mapped while total energies are trained accurately.[5] However, we argue against our results being a case of ad hoc mapping, evidenced by the low standard deviation in $E_{1e}^{I} - E_{1e}^{II}$ values across varying configurations (Figs 6(a),(b)), small total-energy errors for test structures (MAE=2.4 meV/atom), and consistent results from independently trained models (interpolation and extrapolation models). This suggests that the atomic energy mapping shown in Fig. 6 can be generally transferred across various configurations.

In addition, one might question how the lower atomic energy values, compared to the pairwise interactions, can be reconciled with the accurate force predictions shown in Figs. 2,3. We emphasize that the forces are derived from the total energy of the system, not from the atomic energy; i.e., the force is obtained by differentiating the total energy of the system with respect to atomic coordinates, $F_{ia} = dE_{tot}/dr_{ia}$ ($i$ = atomic index, and $a = x, y, z$), and not $F_{ia} = dE_i/dr_{ia}$. As described above, the energy involved in electrostatic interactions between atoms $i$ and $j$ is partially distributed among neighboring atoms. Consequently, the atomic energies of these neighboring atoms incorporate a functional form related to $\mathbf{r}_i$ and $\mathbf{r}_j$, and the correct atomic force can be obtained when differentiating the total energy by $\mathbf{r}_i$ or $\mathbf{r}_j$. That being said, the atomic energy learned by a GNN-IP is hard to be considered as a physical quantity but rather a numerical outcome resulting from the message-passing algorithm. This also provides flexibility in the capability of MLPs to describe non-local interactions, which also helps avoid contradictory scenarios arising from the non-linearity of the Coulomb interaction, as shown in Fig. 1(a).

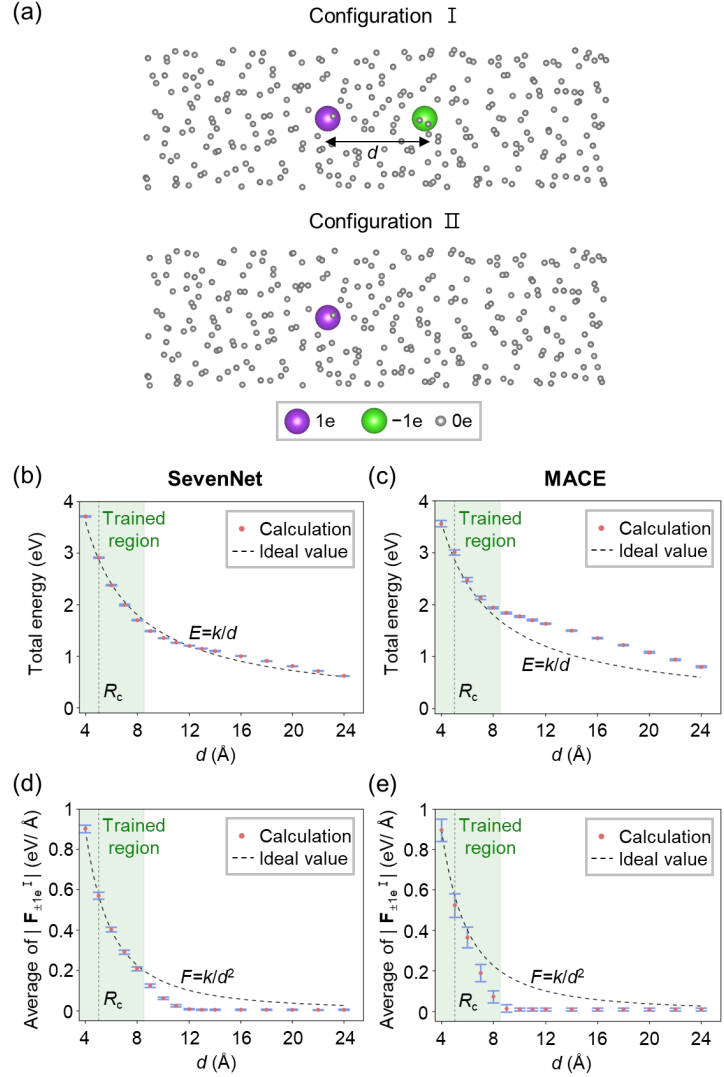

**FIG. 7** (a) Schematic illustration of the toy model with neutral-charge particles. (b,c) Total energy, and (d,e) average absolute forces acting on 1e and −1e particles in configurations I and II calculated by (SevenNet, MACE). Ideal value is plotted as a dotted line each in (b)-(e). Error bars represent the standard deviation of the test results. The trained regions are shaded in green. The vertical dotted line represents the cutoff ($R_c$) of the trained models.

B.3. Toy model with neutral medium particles

To investigate the role of electrostatic interactions in the message-passing algorithm, we perform the same numerical experiment with the system with neutral-charge mediums. Fig. 7(a) illustrates the training set for neutral-center atoms; the configurations for training and testing are identical to those of the charged-particle system, except that the ±0.5e particles are replaced with 0e (neutral) particles. In this setup, the energies of neutral particles are set to 0, and only the electrostatic energies of charged pairs are considered. If a 1e charged particle exists without a corresponding −1e particle (configuration II), the energy is set to 0.

Figs. 7(b),(d) show the calculated energy and force, respectively, obtained from the SevenNet model, and Figs. 7(c),(e) represents the results from the MACE model. Both models are trained with a cutoff of 5 Å using three convolution layers, resulting in an effective cutoff of 15 Å. In the case of SevenNet, we find that the total energies are well predicted in both interpolation and extrapolation regions, in agreement with ref. 19. Interestingly, the total energy prediction of the SevenNet model remains accurate even up to 25 Å (larger than the effective cutoff, 15 Å), indicating that neutral atoms are influenced by charged particles through the unique atomic-energy partitioning scheme of electrostatic energy, as described above. In other words, even when the distance between charged particles exceeds the effective cutoff, a particle positioned between them can contain information from both charged particles. Thus, this intermediate particle can effectively recombine this information, reflecting the energy contribution in the pairwise Coulomb interactions between the charged particles. However, the force values calculated by SevenNet deviate from the ideal values in the extrapolation region, indicating that the extrapolation performance in neutral-center cases is worse compared to charged-particle cases. This suggests that the message-passing algorithm is more effective in systems where physical interactions between charged and medium atoms are present. This could be attributed to the equivariant nature of GNN-IPs: in the neutral system, the forces of neutral atoms are uniformly zero, and differ only based on atomic positions. However, the feature values in the hidden layers must remain non-zero

to enable the recombination of electrostatic interactions via message passing. This requirement can pose challenges for GNN-IP in fitting the data, as the model must satisfy constraints where hidden-layer features are distinct, while ensuring that the derivatives of the total model are zero. Such a requirement imposes restrictions on the model parameters and may lead to inefficiencies in the training process. In contrast, in the charged system, the node features are distinguished by vector quantities (as atomic forces and energies of atoms are different), so the features with $l \geq 1$ can be more flexible for representing this system. Consequently, information transfer is likely more efficient in charged systems than in neutral ones. Further in-depth investigations are necessary in future studies to gain a deeper understanding of this phenomenon.

Meanwhile, the total energy and force predictions made by the MACE model perform less accurately than those of the SevenNet model. Within the interpolation region, the energy predictions from the MACE model agree well with the ideal values but significantly deviate in the extrapolation region. For the force predictions, the MACE model even begins to fail at 7 Å, which falls within the interpolation region. This aligns with the training error values, where the validation errors of SevenNet converge to 0.2 meV/atom and 1.8 meV/Å, while for MACE, they converge to 0.3 meV/atom and 4.1 meV/Å. These results suggest that the basis-degree expansion via the message-passing algorithm in the MACE model is less efficient than the non-linear activation approach of the SevenNet model in a neutral medium. Since the MACE model utilizes a complete basis, its less accurate predictions suggest that a higher degree of basis sets may be necessary for successful learning.

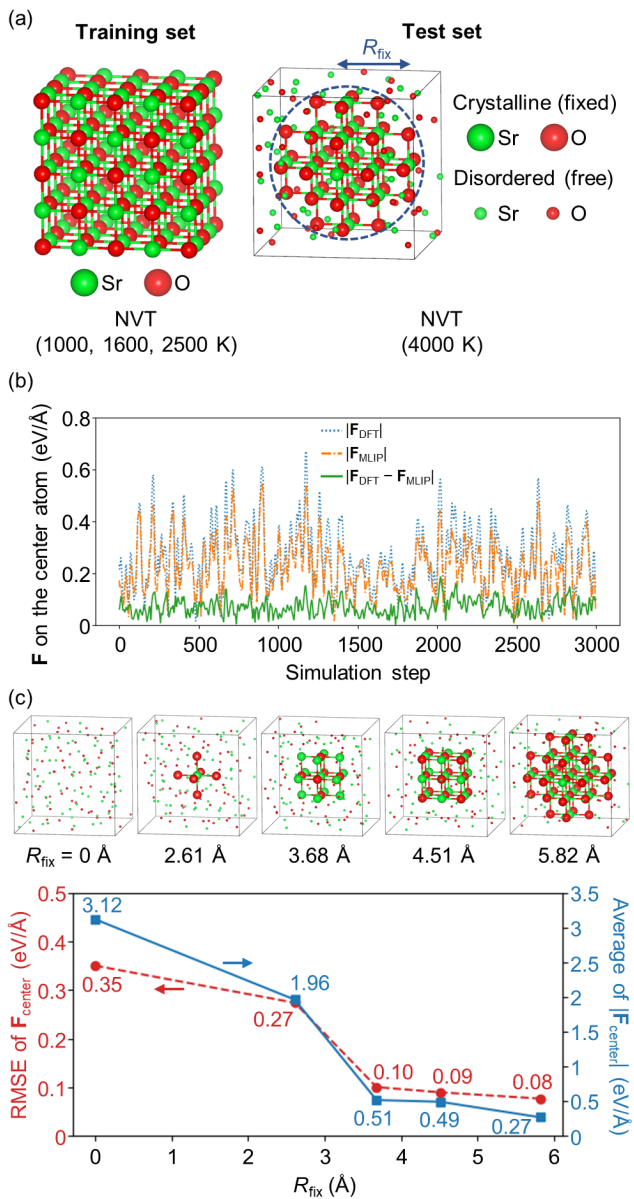

**FIG. 8.** (a) Atomic structures of the training and test sets of the SrO system. In the test structures, the atomic structures inside $R_{fix}$ are fixed during the MD simulations. (b) Force values of the centered atom ($\mathbf{F}_{center}$) calculated by DFT and SevenNet, when $R_{fix}$ = 5.82 Å. (c) Upper: the snapshots of MD simulations with various $R_{fix}$. Lower: RMSE and root mean average of absolute value of $\mathbf{F}_{center}$ as a function of $R_{fix}$.

## C. Learning long-range interactions in DFT data

The previous example demonstrates that the GNN-IPs can learn Coulomb interactions among ideal point charges. However, it remains questionable whether GNN-IPs are capable of capturing non-local interactions in more realistic PES described by DFT, which include a wider range of interactions in addition to Coulomb forces. To test this, we use the model system shown in Fig. 8(a): a SevenNet model is trained on the NVT molecular dynamics (MD) trajectories on the supercell of crystalline SrO. The trained model is tested on the structures where the center part (the atoms closer by $R_{\text{fix}}$ to the center atom) of the cell is fixed to the crystalline structure, and the rest of the part is free to move. The crystalline region is generated by expanding the equilibrium unitcell, and the initial structure of the disorder part is randomly distributed. The structure is equilibrated for 1 ps at the same temperature before the main simulation (see the Methods section for detailed computational settings).

Fig. 8(b) shows the forces calculated by DFT and MLIP on the center atom under the configurations generated with 5.82 Å of $R_{\text{fix}}$. (Note that the center atom refers to a single atom located at the center of the crystalline region, not to all atoms within the crystalline region.) The typical cutoff is 5.5 ~ 6.0 Å, so the force is close to zero when the MLIP is trained with local descriptor-based models. (It is not exactly zero because the cutoff for force extends to twice the cutoff for energies.[33,34] However, it is reasonable to assume that it is close to zero in descriptor-based models, as the electrostatic force decays one order faster than energy with increasing distance.) On the other hand, the force obtained by DFT (blue line) is not zero but oscillates up to ~0.6 eV/Å. The force values calculated by MLIP (orange line) are in good agreement with the DFT (error: green line), indicating that it learns the non-local interactions arising from beyond the cutoff distance. Given that the disordered geometries at the non-local region are not included in the training set but are well predicted by the MLIP, it indicates that the interactions are primarily derived from Coulomb interactions, which the MLIP predicts accurately. Note that the local vibrations included in the training MD trajectories of crystalline structures cause deviations in distance, enabling the MLIP to learn Coulomb interactions and make predictions even for untrained distances, as discussed in the previous subsection. We also evaluate the uncertainty of the atoms in the test set using the standard deviations in the force of ensemble models,[35] as shown in Fig. S2. It can be seen that the uncertainty values are sufficiently low, indicating that the extrapolation is performed effectively across configurations. The uncertainty can be utilized as an indicator for the extent of extrapolation, and its performance for a wider range of error regions will be studied in future works.

Fig. 8(c) illustrates the change of root-mean-square error (RMSE) and the average force values as a function of $R_{\text{fix}}$. While the average absolute force value increases from 5.82 Å to 3.68 Å, the RMSE value remains similar. When $R_{\text{fix}}$ becomes 2.61 Å, both the error value and the absolute force value suddenly increase. This sudden increase supports the conclusion that force values for $R_{\text{fix}}$ larger than 3.68 Å originate from non-local Coulomb interactions, while local interactions become significant for $R_{\text{fix}}$ smaller than 2.61 Å, the region influenced by the trained local configurations.

In the previous example, the charges of the atoms do not significantly vary because SrO is an ionic material. Therefore, it is necessary to verify whether the GNN-IP can accurately capture non-local interactions in systems where the charge values change. To test this, we use a model system consisting of $Au_2$ clusters on MgO and Al-doped MgO supports (Fig. 9(a)), as studied in ref. 9. In this system, the relative energies between wetting and non-wetting configurations change in the case of Al doping, because this alters the charge value of $Au_2$ clusters due to the charge transfer from Al to $Au_2$. In all test simulations, the atomic geometries of the support remain fixed, thus maintaining identical local configurations around the $Au_2$ clusters. Therefore, we rule out the possibility that the MLIP identifies the energy difference from geometric differences, instead of purely inferring the effect of charge transfer from Al to $Au_2$ on the PES of the Au cluster.

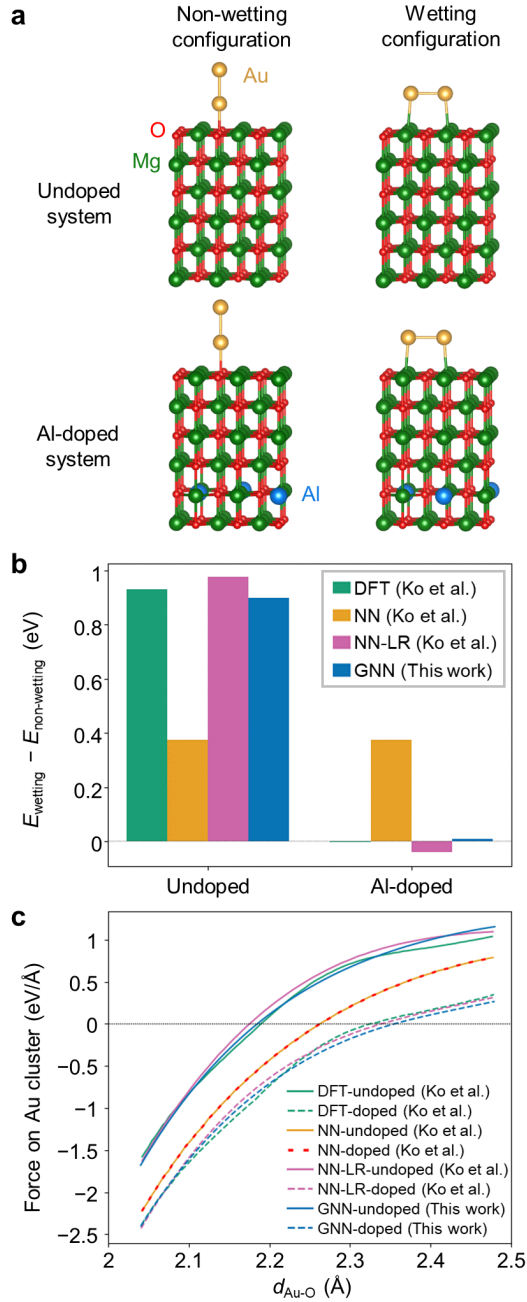

**FIG. 9.** (a) Atomic configurations of Au$_2$ clusters in non-wetting and wetting configurations on undoped and Al-doped MgO, respectively. (b) Energy differences between wetting and non-wetting configurations as calculated by DFT, neural network (NN) potential, NN potential with long-range corrections (NN-LR), and graph neural network (GNN) potential. (c) Sum of forces acting on the Au$_2$ cluster as a function of the distance between lower Au atom and the nearest O atom in the non-wetting configuration. Results of DFT, NN, and NN-LR are extracted from ref. 9.

Fig. 9(b) illustrates the energy differences between the wetting and non-wetting configurations ($E_{\text{wetting}} - E_{\text{non-wetting}}$) for both the MgO and Al-doped MgO systems. The Behler-Parrinello type neural network (NN) potential model[2] yields identical energies for both systems, because it fails to distinguish geometric difference beyond the cutoff. The NN model with long-range corrections (NN-LR) offers a more accurate prediction when compared to DFT results, as detailed in ref. 9. This work finds that the GNN-IP (with cutoff distance of 6 Å) trained with SevenNet model (4 convolution layers) accurately predicts the energy differences between non-wetting and wetting configurations, with the slightly smaller error than NN-LR model.

The PES of the Au cluster is further explored in Fig. 9(c), which depicts the forces acting on the $Au_2$ cluster in a non-wetting configuration. This analysis is conducted while systematically varying the positions of the $Au_2$ clusters, keeping the distance between Au atoms fixed at their equilibrium position. The sum of forces on Au atoms is plotted as a function of the distance between the lower Au atom and the nearest O atom. The trained SevenNet model successfully replicates the PES of the Au clusters on both MgO and Al-doped MgO supports, demonstrating accuracy comparable to the NN-LR model.

We also want to mention that the trained model consists of a 6 Å cutoff with 4 convolution layers. In principle, since the z-coordinate difference between Al and $Au_2$ is about 13 Å, only 3 convolution layers should be sufficient to convey information from Al to the $Au_2$ cluster. However, using this setting, $E_{\text{wetting}} - E_{\text{non-wetting}}$ of the pure and Al-doped cases are 0.721 and 0.141 eV, respectively, which is much less accurate than the case with 4 convolution layers. This indicates that one more layer to the minimum number of layers to convey information is required to accurately describe the interaction, which is also in agreement with the toy model case. This might be attributed to the insufficiency of the message-passing algorithm in fully representing the accurate PES of non-local interactions, as discussed in the previous subsection.

## III. EXTRAPOLATION IN CONFIGURATIONAL SPACE

### A. Theoretical rationalization

In the previous section, we prove that GNN-IPs learn non-local electrostatic interactions accurately enough to extrapolate to untrained configurations. Here, we provide a theoretical explanation for the origin of the extrapolation behavior of GNN-IPs, based on their capability to learn non-local interactions combined with their embedding characteristics.

In ref. 5, it is demonstrated that the atomic energy learned from MLIPs ($E_{at,i}$, $i$: atomic index) can be divided into three terms; kinetic energy ($E_{kin,i}$), exchange-correlation energy ($E_{XC,i}$), and Coulomb energy ($E_{Coul,i}$):

$$E_{at,i} = E_{kin,i} + E_{XC,i} + E_{Coul,i}. \tag{3}$$

These terms are defined as follows within the semilocal density approximation:

$$E_{kin,i} = -\frac{1}{2}\int \nabla_{\mathbf{r}}^2 \rho_{at,i}(\mathbf{r},\mathbf{r}')|_{\mathbf{r}=\mathbf{r}'}d\mathbf{r}', \tag{4}$$

$$E_{XC,i} = \int \rho_i(\mathbf{r})\varepsilon_{XC}(\rho_i(\mathbf{r}),\nabla\rho_i(\mathbf{r}))d\mathbf{r}, \tag{5}$$

$$E_{Coul,i} = \frac{1}{2}\sum_{j\neq i}\frac{\rho_{tot,i}(\mathbf{r})\rho_{tot,j}(\mathbf{r}')}{|\mathbf{r}-\mathbf{r}'|}d\mathbf{r}d\mathbf{r}' + \frac{1}{2}\int\frac{\rho_i(\mathbf{r})\rho_i(\mathbf{r}')}{|\mathbf{r}-\mathbf{r}'|}d\mathbf{r}d\mathbf{r}' - \frac{1}{2}\int\frac{q_i\rho_i(\mathbf{r})}{|\mathbf{r}-\mathbf{r}_i|}d\mathbf{r}, \tag{6}$$

where, $\rho_i$ represents the electron density within the partitioned cells (e.g. Voronoi tessellation) around atom $i$. (Note that there is no way to determine the specific scheme the MLIP model uses to map atomic energies, as it often varies depending on the details of the training set and hyperparameters.)[5] The atomic density matrix ($\rho_{at,i}$) and total charge density ($\rho_{tot,i}$) are defined as follows:

$$\rho_{\text{at},i}(\mathbf{r},\mathbf{r}') = \rho_{ii}(\mathbf{r},\mathbf{r}') + \frac{1}{2}\sum_{\substack{j\neq i}}^{r_{ij}<R_c^{\text{kin}}} \rho_{ij}(\mathbf{r},\mathbf{r}'), \quad (7)$$

$$\rho_{\text{tot},i}(\mathbf{r}) = q_i\delta(\mathbf{r}-\mathbf{r}') - \rho_i(\mathbf{r}). \quad (8)$$

$R_c^{\text{kin}}$ is the cutoff distance for kinetic energy. In the case of Coulomb interactions, in the previous section, we show that atomic energy learned from GNN-IPs does not only include the sum of pairwise interactions involving *i*, but also includes a fraction of interactions between atoms around *i* (Fig. 3). Therefore, $E_{\text{Coul},i}$ can be expressed as follows:

$$E_{\text{Coul},i} = \frac{1}{2}\sum_{\substack{k\neq j}}^{r_{ij},r_{ik}<R_c^{\text{Coul}}} w_{\{n_{ijk}\}} \frac{\rho_{\text{tot},j}(\mathbf{r})\rho_{\text{tot},k}(\mathbf{r}')}{|\mathbf{r}-\mathbf{r}'|} d\mathbf{r}d\mathbf{r}' + \frac{1}{2}\int \frac{\rho_i(\mathbf{r})\rho_i(\mathbf{r}')}{|\mathbf{r}-\mathbf{r}'|} d\mathbf{r}d\mathbf{r}'$$

$$-\frac{1}{2}\int \frac{q_i\rho_i(\mathbf{r})}{|\mathbf{r}-\mathbf{r}_i|} d\mathbf{r}, \quad (9)$$

where $R_c^{\text{Coul}}$ is the cutoff distance for a Coulomb interaction and $w_{n_{ijk}}$ is the contribution of the electrostatic energy between *k* and *j* to the atomic energy of *i*. $\{n_{ijk}\}$ represents the neighboring atoms around *i*, *j*, and *k*, as these atoms can also contribute to the electrostatic energy term, as discussed in the previous section on atomic energy mapping. The sum of $w_{n_{ijk}}$ over index *i* is 1.

Since $E_{\text{XC},i}$ is defined locally in the semilocal density approximation and $E_{\text{kin},i}$ often decays within the a reasonable cutoff distance (further discussed in the next subsection), these terms can be classified as local terms. We further divide $E_{\text{Coul},i}$ into local and non-local terms:

$$E_{\text{Coul},i} = E_{\text{Coul},i}^{\text{non-local}} + E_{\text{Coul},i}^{\text{local}}, \quad (10)$$

where each term is expressed as follows:

$$E_{\text{Coul},i}^{\text{local}} = \frac{1}{2}\int \frac{\rho_i(\mathbf{r})\rho_i(\mathbf{r}')}{|\mathbf{r}-\mathbf{r}'|} d\mathbf{r}d\mathbf{r}' - \frac{1}{2}\int \frac{q_i\rho_i(\mathbf{r})}{|\mathbf{r}-\mathbf{r}_i|} d\mathbf{r}, \quad (11)$$

$$E_{\text{Coul},i}^{\text{non-local}} = \frac{1}{2} \sum_{k \neq j}^{\mathbf{r}_{ij}, \mathbf{r}_{ik} < R_c^{\text{Coul}}} w_{\{n_{ijk}\}} \frac{\rho_{\text{tot},k}(\mathbf{r}) \rho_{\text{tot},j}(\mathbf{r}')}{|\mathbf{r} - \mathbf{r}'|} d\mathbf{r} d\mathbf{r}'. \tag{12}$$

$E_{\text{Coul},i}^{\text{non-local}}$ converges to a simple point-charge interaction term, i.e., $kQ_{1,\text{tot}}Q_{2,\text{tot}}/r_{12}$ (where $Q_{1,\text{tot}}$ and $Q_{2,\text{tot}}$ is the total charges of atoms 1 and 2), as the distance between atoms increases regardless of the specific charge distributions of atoms, which can be derived from the far-field approximation with multipole expansion (see the Supplementary Material for details). Because of this universal form of the Coulomb interaction in a non-local region, if a GNN-IP model is capable of predicting total charge values of atoms, this term can be accurately learned from a reasonable dataset, such as distorted crystal structures (NVT MD), as shown in the previous section.

We expect that the short-range terms, $E_{\text{Coul},i}^{\text{local}}$, $E_{\text{kin},i}$, and $E_{\text{XC},i}$ are difficult to extrapolate to untrained configurations due to the complexity of quantum mechanical PES. Even so, the GNN-IPs utilizes embedding methods, so the knowledge of local structures learned from one composition can be transferred to other compositions. This is because all elements share the same networks and differ only by the parts related to trainable embedding vectors. Therefore, if the GNN-IPs are trained on a large database, they would learn the local energy terms of trained configurations but in untrained compositions.

Overall, the short-range terms are interpolated in the configuration space and extrapolated in the compositional space and the non-local terms are extrapolated in the configuration space. This indicates that GNN-IPs can describe untrained structures if their local structures are included in the training data. This explains how GNN-IPs accurately describe untrained structures.

There still exists debate on the extent to which distance is considered the transferable local region, which varies across specific systems. Therefore, if the lower limit of the distance at which the Coulomb interactions are approximated by the point-charge model is below the transferable distance for short-range interactions, extrapolation can be performed accurately. Conversely, if there exists a distance range that is longer than the local region but shorter than the point-charge approximation region, Coulomb interactions in this intermediate region becomes difficult to extrapolate using a GNN-IP model. For instance, in the case of SrO crystal/disorder system, this transferable local distance is expected to lie between 2.61 Å and 3.68 Å, as shown in Fig. 8(c), because the error increases significantly beyond these distances. For the long-range part, if we assume a Gaussian shape charge distribution (with the width being the covalent radius) among atoms,[9] the Coulomb interaction between Sr and O becomes $0.91kQ_{1,tot}Q_{2,tot}/r_{12}$ when $r_{12}$ is 3.48 Å, and $0.99kQ_{1,tot}Q_{2,tot}/r_{12}$ at 5.22 Å. These distance limits are similar for local and non-local parts for the SrO crystal/disorder system, which explains why the successful extrapolation in Fig. 8(c) is made above $R_{fix}$ of 3.68 Å.

**B. Numerical experiments with UIPs**

In UIP models such as MACE-MP-0 and SevenNet-0, non-local Coulomb interactions can be effectively extrapolated to diverse materials by learning the relaxation trajectories of crystal structures in the Materials Project database. We perform numerical experiments with SevenNet-0 to confirm this hypothesis. Fig. 10(a) shows the force values of the centered atom computed by DFT and SevenNet-0, and the RMSE values of the SrO crystal/disorder simulation trajectory, as described in Fig. 8(a). The average RMSE is 0.119 eV/Å, which is slightly larger than the value shown in Fig. 8(c) (0.076 eV/Å). Even so, the overall trend in SevenNet-0 values systematically aligns with the reference DFT data. Therefore, this indicates that SevenNet-0 learned the non-local Coulomb interaction with reasonable accuracy.

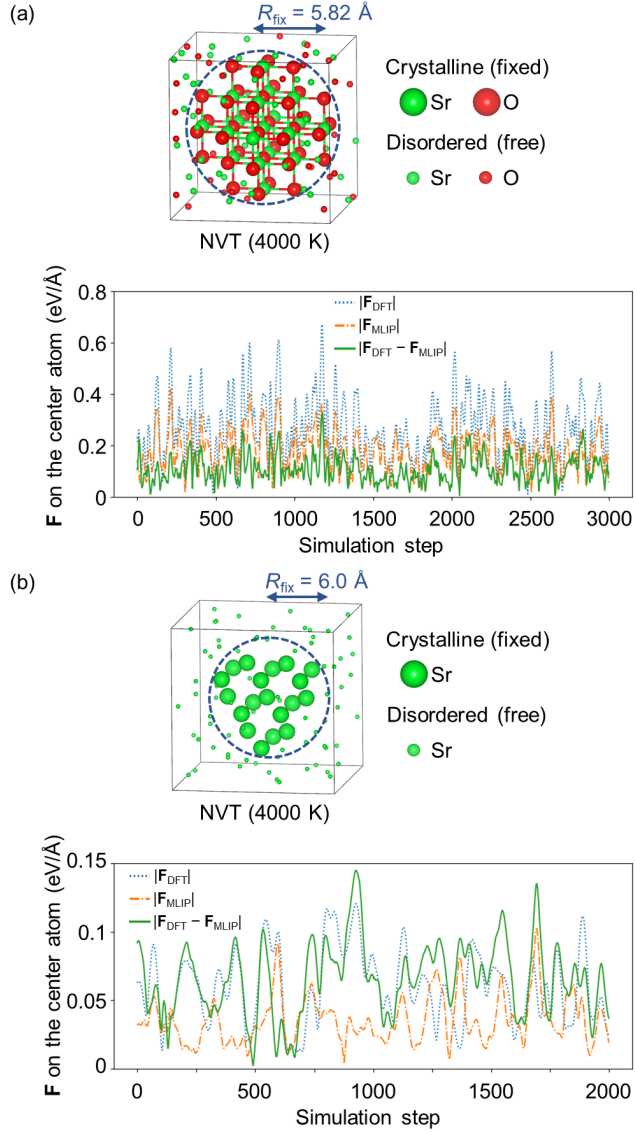

**FIG. 10.** Atomic structures and force values of the centered atom calculated by DFT and SevenNet-0 for (a) SrO and (b) Sr. In the test structures, the atomic structures inside $R_{fix}$ are fixed during the MD simulations.

We also assess the contribution of the kinetic energy term by utilizing the Sr system, as shown in Fig. 10(b). This material is unary and metallic, so we can assume that the electrostatic interaction beyond the cutoff is mostly screened, and the force applied to the center atom corresponds to the kinetic energy term. The graph in Fig. 10(b) shows that the force error has an average value of 0.066 eV/Å. SevenNet-0 does not accurately predict the overall trend, indicating that the kinetic energy contribution is not extrapolated. Even so, the absolute error values are less than the typical force resolution of DFT calculations, 0.1 eV/Å, so we expect that this will not introduce significant errors in typical simulations.

Finally, it is important to discuss the limitations of the distance range that permits effective extrapolation. Based on the above discussion with Gaussian-shaped charges, the Coulomb interaction is expected to be well-approximated beyond 3–5 Å. Distances below this range correspond to the region that can be learned through interpolation in compositional space. The kinetic energy is also expected to interpolate within this local distance range. However, determining the exact limit of interpolation distance for kinetic energy is challenging, as the kinetic energy term is hard to be approximated to a simple functional approximation like the Coulomb interaction. Since the kinetic energy term is likely a major source of extrapolation error, it is crucial to identify effective metrics and distance ranges for interpolating this term in future studies.

## IV. DISCUSSION

### A. Performance of UIPs

According to the theory suggested here, increasing the number of layers enhances the ability to extrapolate non-local electrostatic interactions, thereby affects the extrapolation capability. This might provide insights into the relative performance between SevenNet-0[25] and MACE-MP-0,[24] the top two open-source models in the Matbench discovery benchmark platform[36] (https://matbench-discovery.materialsproject.org/, accessed 2024-07-30). SevenNet-0 shows better performance than MACE-MP-0, even though SevenNet-0 contains fewer parameters (842.4K) compared to MACE-MP-0 (4.7M) and uses the same size of training data (146K). This likely originates from the larger number of convolution layers in SevenNet-0 (five layers) compared to MACE-MP-0 (two layers), suggesting that the number of convolution layers is an important factor in the extrapolation capability. This is further supported from the fact that MACE-MP-0 even uses a higher-level message-passing algorithm compared to SevenNet-0, including many-body order message terms (in MACE-MP-0, up to 4-body terms), and utilizes a higher order of spherical expansion ($l_{max}$ = 3) compared to SevenNet-0 ($l_{max}$ = 2).

Specifically, in the previous section, we demonstrate that the message-passing algorithm does not provide full accuracy in extrapolating electrostatic interactions, particularly in the last convolution layer. MACE-MP-0, with only two convolution layers and a 6 Å cutoff, has a lower effective cutoff for accurately extrapolating electrostatic interactions than 12 Å. In contrast, SevenNet-0 includes five convolution layers with a 5 Å cutoff, allowing for extrapolation of electrostatic interactions over a distance of at least up to 20 Å. Thus, the effective cutoff radius for accurate electrostatic interactions is approximately twice as large in SevenNet-0 compared to MACE-MP-0. This difference means that MACE-MP-0 must train with larger noise from electrostatic energies originating beyond its effective cutoff distance. As suggested in refs. 5,6, this leads to a random allocation of error terms, which in turn reduces accuracy, particularly when encountering untrained compositions and structures. We also note that the difference in self-connection methods can affect extrapolation performance, as suggested in ref. 25.

Here, we focus solely on the non-local interaction term learned from the message-passing algorithm and do not specifically delve into the extrapolation capability of local terms. We anticipate that local terms are interpolated in the configuration space by embedding characteristics, although the precise interpolation ability of GNN-IP models remains unclear. Therefore, systematic methods for numerical experiments and analysis techniques, such as Uniform Manifold Approximation Projection (UMAP)[37] as used in ref. 24, would need to be further developed.

**B. Fundamental error in local MLIPs**

We want to mention the typical training error values in local descriptor-based models. As shown in Figs. 8,10, the magnitudes of the non-local interaction terms (0.27 eV/Å in average for SrO case) can be larger than the typical error range of DFT calculations, 0.1 eV/Å. The typical force error values in descriptor-based models are about 0.1 to 0.3 eV/Å (for instance, refs. 38–40), and the error can even reach about 0.5 eV/Å when dealing with ionic disordered structures (for instance, refs. 41,42). Large parts of these error values correspond to a fundamental limit that cannot be further reduced, rather than errors from the insufficient fitting of machine learning models, given the non-local forces in DFT calculations identified in this study. Note that the error values can be less than 0.1 eV/Å when the structure is not significantly deviated from the equilibrium crystal structures, even if they contain ionic characteristics (for instance, refs. 43,44). Systematic investigations into the fundamental errors arising from non-local interactions in more diverse material groups will be further studied in future works. Unlike forces, the energy errors are often minimized to a few meV/atom, which is smaller than the typical DFT error, 10 meV/atom. This might be attributed to the electrostatic interactions being averaged out and arbitrarily allocated to the elements, as confirmed in refs. 5,6.

**C. Applicability of GNN-IPs in diverse applications**

We suggest that GNN-IP models can be efficient tools for investigating systems that exhibit non-local interactions due to insufficient screening, even without explicitly implementing Coulomb interaction terms (for instance, see ref. 9). However, the following considerations must be addressed for further practice: (1) The message-passing algorithm in GNN-IP models, which requires a third atom connecting two others beyond the cutoff, generally functions well in condensed matter systems where atoms are closely interconnected. Nevertheless, in sparser systems like molecular liquids, GNN-IP may encounter challenges in training the PES. (2) Technically, the evaluation of electrostatic energy in periodic systems necessitates calculations of an infinite sum across all periodic images, typically using methods like Ewald summation. Currently, conventional GNN-IP models are not equipped to perform this calculation. Long-range electrostatic effects (greater than 20 Å) are generally diminished due to effective screening in most condensed matter systems, and the message-passing algorithm further reduces the remaining errors in the medium-range (10–20 Å) regions as shown in this work. However, careful attention is still required to accurately account for factors such as dielectric screening and the specific charge distribution of each system. As an example, we suggest that electrochemical environments, such as interfaces of catalysts in aqueous solvents and Li-ion batteries, can be target systems studied with GNN-IPs, as they have larger dielectric constants that effectively screen out long-range interactions.

**D. Long-range interactions other than Coulomb forces**

In this study, we only discuss the long-range force associated with Coulomb forces. However, there still exists the possibility that other long-range forces affect the extrapolation capability of GNN-IPs. Besides Coulomb interactions, the most common long-range interaction is the van der Waals (vdW) interaction. VdW interactions depend on $\sim 1/r^6$ (where $r$ is the interatomic distance), which decays much faster than the Coulomb interaction ($1/r$). However, as vdW forces mostly involve attractive terms, their cumulative effect can be significant, particularly in systems with heavy atoms, where they contribute to long-range forces.

Most of large GNN-IPs, such as SevenNet-0 and MACE-MP-0, are trained on data obtained from the PBE functional, so these interactions do not impact the extrapolation capabilities of these GNN-IPs. To include vdW effects in semilocal DFT calculations, vdW interactions are often approximated using analytic forms such as D2 and D3 functionals,[45,46] which can be applied without quantum mechanical calculations. Therefore, they can be directly incorporated into MLIPs (for instance, see refs. 47,48). These functionals are simple and broadly applicable across diverse systems without requiring parameter modifications for each system. In some cases, however, vdW interactions cannot be described using simple analytic forms, requiring high-accuracy DFT calculations, such as MBD@rsSCS[49] and random-phase approximation (RPA).[50] In these cases, the interactions must be trained on DFT data that explicitly include vdW effects. We also note that magnetic interactions, such as RKKY interactions, can be long-range, as they depend on $\sim 1/r^3$. Since these interactions decay faster than Coulomb interactions, they are expected to be adequately addressed within the effective cutoff of conventional GNN-IP models. However, it remains unclear whether these interactions can be extrapolated to untrained domains. This remains an open area for future exploration.

## V. CONCLUSION

In summary, we provide a theoretical explanation for how GNN-IPs accurately predict untrained configurations. First, we prove that GNN-IPs learn non-local electrostatic interactions via the message-passing algorithm. This is tested with toy models consisting of point charges, which verifies that GNN-IPs accurately predict Coulomb interactions over a range of distances not included in the training set. Next, we verify that electrostatic interactions can be inferred from distorted crystal structures generated by MD simulations, allowing for extrapolation to electrostatic interactions contributed by disordered structures. We also verify that electrostatic interactions can be learned in cases where charge values change due to charge transfer, by using the model system where $Au_2$ is adsorbed on undoped/doped MgO slabs. Finally, we demonstrate that the capability of learning electrostatic interactions and the embedding characteristics of GNN-IPs are the origin of their extrapolation capability. Based on this, we suggest that the number of convolution layers is a critical factor for enhancing the extrapolation capability of UIPs.

## VI. Methods

### A. Toy model

The lattice dimensions of each cell are set to $10\times10\times(24+d)$ Å$^3$, configured in a tetrahedral shape. Each cell contains one 1e and one −1e particle, where the coordinates of these particles are set at (5, 5, 12) and (5,5,12+$d$), respectively, along with 120+5$d$ particles each of 0.5e and −0.5e, which are randomly distributed within the simulation cell. The minimum interatomic distance cutoff for the random distribution is set at 1.5 Å. To eliminate periodic effects, 6 Å of vacuum layers are inserted in all directions, which is larger than the cutoff of 5 Å. In total, 200 structures are constructed for each $d$ for training set. The test set is generated by the same method as for the training set.

### B. SrO, Sr crystal/disorder system

All DFT calculations are performed with the VASP code,[51] using projector augmented-wave (PAW) pseudopotentials[52] and the PBE exchange−correlation functional.[53] For the SrO system, the training set includes computation results for a $10.40\times10.40\times10.40$ Å$^3$ simulation cell containing 32 Sr and 32 O atoms. The test cell consists of a $13.63\times13.63\times13.63$ Å$^3$ simulation cell with 72 Sr and 72 O atoms. NVT MD simulations are performed for 10 ps each at 1000 K, 1600 K, and 2500 K, with the training set sampled every 20 fs. For the test structures, initial test structures are generated by placing fixed crystal compositions in the middle of the cell and randomly positioning the rest of the atoms in the remaining space. MD simulations for the test set are performed for 6 ps at 4000 K. For the Sr system, the test set includes computation results for a 17.29×17.29×17.29 Å$^3$ simulation cell containing 96 Sr atoms, with the density set identical to the Sr crystal. The simulations are performed for 5 ps at 4000 K. The cutoff energy values for SrO and Sr are set to 500 eV and 200 eV, respectively, as determined by the convergence test. A single gamma-point and a (1/4, 1/4, 1/4) **k**-point are used for training and test simulations, respectively, in the case of SrO. A single (1/4, 1/4, 1/4) **k**-point is used for Sr.

We note that the cutoff energies and **k**-point densities used in our simulations are less accurate than those used in the Materials Project: 520 eV for the cutoff energy, and 4×4×4 and 6×6×6 gamma-centered **k**-point meshes for SrO and Sr unit cells, respectively. Since SevenNet-0 is trained with the Materials Project database, this discrepancy might cause errors in the comparisons made in Fig. 7. To check this, we perform tests using snapshots extracted from the test simulation trajectories and find that our settings cause deviations of only 0.012 eV/Å and 0.004 eV/Å for SrO and Sr, respectively, compared with the Materials Project settings (**k**-point meshes are reduced to 2 × 2 × 2, and the cutoff is set to 520 eV).

## C. Au-MgO and Au/Al-MgO system

We employ the same training set and test system as utilized in ref. 9. For the training set, it is stated that random displacements are applied to the configurations in the trajectory shown in Fig. 6(c), with a standard deviation of 0.02 Å for the substrate atoms and 0.1 Å for the gold atoms. Half of the training set corresponds to the undoped substrate, and the other half corresponds to the doped substrate. For each substrate configuration, half of the samples were generated with the $Au_2$ cluster in its wetting configuration, while the remaining half featured the cluster in its non-wetting configuration.

### D. Machine-learned interatomic potentials

The SevenNet code[25] is used with the NequIP architecture,[15] and the simulations are performed with the LAMMPS code.[54] The model designed for the point-charge system comprises 5 convolution layers, with a maximum rotation order ($l_{max}$) of 3 and 32 features for each rotation order. The model for the SrO system includes 3 convolution layers, with a $l_{max}$ of 3. The model for the Au-MgO system includes 4 convolution layers, with a $l_{max}$ of 2 and 16 features for each rotation order. The cutoff radius of the toy model, SrO system, and Au-MgO system are set to 5, 5, and 6 Å, respectively. For all models, the invariant radial networks utilize a trainable Bessel basis of size 8 and consist of two hidden layers with 64 nodes each, featuring SiLU nonlinearities between the layers. The train/validation sets are split by the ratio of 9:1. For the MACE code, we use the similar settings as in SevenNet, except for using a correlation order of 3. All input files are publicly available at https://doi.org/10.6084/m9.figshare.26496160.v2.

### E. Visualization of atomic structures

All atomic configurations in this paper is drawn with VESTA code.[55]

## Supplementary Material

See the supplementary material for detailed descriptions of far-field approximation, a single-layer GNN-IP model, and uncertainties in the SrO system.

# Data availability

The codes for training MLIPs, SevenNet and MACE, are open-source, and the links can be found in the original papers.[14,25] The training sets, test sets, input files, and the trained models are provided at https://doi.org/10.6084/m9.figshare.26496160.v6.[56] The training set for the Au-MgO system is available in the repository distributed by the previous study of Ko et al.[57]

# Acknowledgements

This work was supported by the Nano & Material Technology Development Programs through the National Research Foundation of Korea (NRF) funded by Ministry of Science and ICT (RS-2024-00407995 & RS-2024-00450102).

# Competing interests

The author declares no competing interests.